\documentclass[twocolumn]{aastex61}
\received{May 23, 2017}
\revised{Aug 22, 2017}
\accepted{Sep 16, 2017}
\submitjournal{ApJ}

\shorttitle{A Face-on Accretion System in High-Mass Star-Formation}
\shortauthors{Motogi et al.}

\begin{document}

\title{A Face-on Accretion System in High-Mass Star-Formation: \\
Possible Dusty Infall Streams within 100 AU}

\correspondingauthor{Kazuhito Motogi}
\email{kmotogi@yamaguchi-u.ac.jp}

\author{Kazuhito Motogi }
\affil{Graduate School of Sciences and Technology for Innovation,Yamaguchi University, Yoshida 1677-1, Yamaguchi 753-8512, Japan}

\author{Tomoya Hirota}
\affil{Mizusawa VLBI Observatory, National Astronomical Observatory of Japan, Osawa 2-21-1, Mitaka, Tokyo 181-8588, Japan}
\affil{Department of Astronomical Sciences, SOKENDAI (The Graduate University for Advanced Studies), Osawa 2-21-1, Mitaka, Tokyo 181-8588, Japan}

\author{Kazuo Sorai}
\affil{Department of Physics, Faculty of Science, Hokkaido University, Sapporo 060-0810, Japan}
\affil{Department of Cosmosciences, Graduate School of Science, Hokkaido University, Sapporo 060-0810, Japan}

\author{Yoshinori Yonekura}
\affil{Center for Astronomy, Ibaraki University, 2-1-1 Bunkyo, Mito, Ibaraki 310-8512, Japan}

\author{Koichiro Sugiyama}
\affil{Mizusawa VLBI Observatory, National Astronomical Observatory of Japan, Osawa 2-21-1, Mitaka, Tokyo 181-8588, Japan}

\author{Mareki Honma}
\affil{Mizusawa VLBI Observatory, National Astronomical Observatory of Japan, Hoshigaoka 2-12, Mizusawa, Oshu, Iwate 023-0861, Japan}
\affil{Department of Astronomical Sciences, SOKENDAI (The Graduate University for Advanced Studies), Osawa 2-21-1, Mitaka, Tokyo 181-8588, Japan}

\author{Kotaro Niinuma}
\affil{Graduate School of Sciences and Technology for Innovation,Yamaguchi University, Yoshida 1677-1, Yamaguchi 753-8512, Japan}

\author{Kazuya Hachisuka}
\affil{Mizusawa VLBI Observatory, National Astronomical Observatory of Japan, Hoshigaoka 2-12, Mizusawa, Oshu, Iwate 023-0861, Japan}

\author{Kenta Fujisawa}
\affil{The Research Institute for Time Studies, Yamaguchi University, Yoshida 1677-1, Yamaguchi 753-8511, Japan}

\author{Andrew J. Walsh}
\affil{International Centre for Radio Astronomy Research, Curtin University, Bentley, WA 6102, Australia}

\begin{abstract}

We report on interferometric observations of a face-on accretion system around the High-Mass young stellar object, G353.273+0.641. 
The innermost accretion system of 100 au radius was resolved in a 45 GHz continuum image taken with the Jansky-Very Large Array. 
Our spectral energy distribution analysis indicated that the continuum could be explained by optically thick dust emission. 
The total mass of the dusty system is $\sim$ 0.2 $M_{\sun}$ at minimum and up to a few $M_{\sun}$ depending on the dust parameters. 
6.7 GHz CH$_{3}$OH masers associated with the same system were also observed with the Australia Telescope Compact Array. 
The masers showed a spiral-like, non-axisymmetric distribution with a systematic velocity gradient. 
The line-of-sight velocity field is explained by an infall motion along a parabolic streamline that falls onto the equatorial plane of the face-on system. 
The streamline is quasi-radial and reaches the equatorial plane at a radius of 16 au. 
This is clearly smaller than that of typical accretion disks in High-Mass star formation, indicating that the initial angular momentum was very small, 
or the CH$_{3}$OH masers selectively trace accreting material that has small angular momentum. 
In the former case, the initial specific angular momentum is estimated to be 8 $\times$ 10$^{20}$ ($M_{*}$$/$10 $M_{\sun}$)$^{0.5}$ cm$^{2}$ s$^{-1}$, 
or a significant fraction of the initial angular momentum was removed outside of 100 au. 
The physical origin of such a streamline is still an open question and will be constrained by the higher-resolution ($\sim$ 10 mas) thermal continuum and line observations with ALMA long baselines. 

\end{abstract}


\section{Introduction} \label{sec:intro}
A dozen or more disk-like structures in High-Mass star-formation have been reported in this decade \citep[e.g., ][and references therein]{2006Natur.444..703C, 2016A&ARv..24....6B, 2017A&A...602A..59C}. 
Some of these sources probably trace circumcluster envelopes  \citep[so-called "toroid"; ][]{2005A&A...435..901B, 2011A&A...525A.151B} rather than individual accretion disks. 
Such a structure suggests a hierarchical fragmentation and accretion process around a High-Mass young stellar object (HMYSO). 
The innermost compact disks ($<$ 100 au) were resolved in some very high-resolution observations \citep[e.g.,][]{2010Natur.466..339K, 2014ApJ...782L..28H, 2016ApJ...833..238H, 2017NatAs...1E.146H, 2016ApJ...833..219P}.  

Detailed theoretical studies have suggested that the evolution of a High-Mass protostar depends strongly on the accretion rate ($\dot{M}_{\rm acc}$) onto the protostellar surface and the exact accretion geometry 
 \citep{2009ApJ...691..823H, 2010ApJ...721..478H}. 
For example, a high accretion rate ($>$ 10$^{-4}$ M$_{\sun}$ yr$^{-1}$), which is expected for High-Mass star-formation even in disk accretion \citep[e.g., ][and references therein]{2007ARA&A..45..481Z}, greatly swells protostars, resulting in much lower stellar temperatures. 
This significantly delays stellar ignition (up to 20 $M_{\sun}$), changing the history of the energy feedback (e.g., bolometric luminosity, total amount of $UV$-photons, etc.). 
\citet{2011MNRAS.416..972D} have reported that such an effect can consistently explain the observed luminosity function of HMYSOs. 
Observational studies of the innermost accretion properties will thus be an essential task in the next decade. 

Almost all of the known circumstellar systems in High-Mass star-formation are in a nearly edge-on orientation \citep[e.g.,][]{2005Natur.437..109P, 2007ApJ...666L..37T, 2010ApJ...708...80M, 2012ApJ...752L..29C, 2013A&A...552L..10S, 2014A&A...571A..52B, 2014A&A...566A..73C, 2016ApJ...833..238H, 2017NatAs...1E.146H, 2016ApJ...833..219P}. 
This bias is simply due to their easy identification even with limited angular resolution, based on the velocity gradient of rotating disks and/or envelopes. 
Such edge-on sources are suitable for kinematic studies; however, the innermost region is significantly obscured by a surrounding envelope and an edge-on accretion disk. 
It is also difficult to study detailed structures apart from the rather conservative picture of a homogeneous rotating and/or infalling disk, even with the extreme sensitivity and resolution of the Atacama Large Millimeter/Submillimeter Array (ALMA). 
A face-on accretion system is, in contrast, a more preferable target for inner-disk observations, 
where the self-shielding effect is minimized and an outflow cavity can reduce the total optical depth along the line of sight (LOS). 
In addition to the observational feasibility, one can directly observe radial structures such as surface density, temperature, and chemical composition, etc. 
Such information is essential for understanding the angular momentum transfer and heating/cooling processes in the innermost region. 
The latter, in particular, controls the degree of fragmentation and the maximum stellar mass \citep[e.g., ][]{2010ApJ...713.1120K, 2011ApJ...740..107C}. 

In this paper, we report on interferometric observations of a nearly face-on HMYSO. 
The target source, G353.273+0.641 (hereafter G353), is a relatively nearby \citep[1.7 kpc;][]{1978A&A....69...51N, 2016PASJ...68...69M} HMYSO in the southern sky. 
The mass of the HMYSO is still uncertain; however, association with class II CH$_{3}$OH maser emission \citep[e.g.,][]{2008MNRAS.386.1521C}, which exclusively traces HMYSOs \citep[e.g.,][]{2003A&A...403.1095M}, suggests that the source is already a High-Mass object. 
G353 is also known as a dominant blueshifted maser (DBSM) source that is a class of 22 GHz H$_{2}$O masers showing a highly blue-shift dominated spectrum \citep{2008MNRAS.386.1521C}. 
The LOS velocities typically range from -120 to -45 km s$^{-1}$ and the systemic velocity ($V_{\rm sys}$) is $\sim$ -5.0 km s$^{-1}$ in the case of G353 \citep[e.g.,][]{2008MNRAS.386.1521C, 2011MNRAS.417..238M, 2016PASJ...68...69M}. 

\citet{2008MNRAS.386.1521C} proposed that DBSMs are candidates of HMYSOs with a face-on protostellar jet. 
\citet{2013MNRAS.428..349M} have actually found a faint radio jet and an extremely high-velocity thermal SiO jet in G353. 
High-resolution observations using the VLBI Exploration of Radio Astrometry (VERA) have shown that 
the H$_{2}$O maser also traces a very compact bipolar jet within 400 au around the radio continuum peak \citep{2011MNRAS.417..238M, 2016PASJ...68...69M}.  
The inclination angle of the jet, which was measured by maser proper motions, is 8$\degr$  -- 17$\degr$ from the LOS. 
In addition, \citet{2016PASJ...68...69M} have found that the jet is recurrently ejected, and also accelerated within 100 au from the host HMYSO. 
The lack of any OH maser emission also indicates that the host HMYSO is likely to be in an early phase of evolution \citep{2010MNRAS.406.1487B}. 
These facts confirmed that G353 is, at present, the best candidate of a face-on HMYSO in the active accretion phase. 

We have searched for a face-on accretion system associated with G353, 
via the highest-resolution imaging of 45 GHz continuum emission using the Jansky-Very Large Array (J-VLA). 
We also performed mapping observations of the 6.7 GHz class II CH$_{3}$OH maser, and also 6/9-GHz band (C/X-band)  continuum by the Australia Telescope Compact Array (ATCA). 
This maser is a plausible tracer of an accretion disk and/or envelope in High-Mass star-formation \citep[e.g., ][]{2010A&A...517A..71S, 2011A&A...536A..38M, 2014A&A...566A.150M, 2014A&A...562A..82S}. 
In addition to these, we investigated a bolometric luminosity of G353 using some archival infrared (IR) data, in order to estimate a mass of the host HMYSO. 

\section{Observations and data reductions} \label{sec:obs}
The ATCA and J-VLA observations are detailed in the following sub-sections. 
Table \ref{tbl-1} presents brief summaries. 
The phase-tracking center of G353 was set to 17$^{\rm h}$26$^{\rm m}$01$^{\rm s}$.59, -34\degr15\arcmin14\arcsec.90 (J2000.0) in both observations.

\subsection{J-VLA observations} \label{sec:obs:vla}
Two successive observing blocks were allocated on 2014 February 18. 
There were 24 and 26 available antennas for the first and second observing blocks, respectively. 
The total observing time was 2.5 hr at an elevation above 20$\degr$. 
Observations were performed with the WIDAR correlator \citep{2011ApJ...739L...1P} in 8-bit sampling, dual polarization mode. 
Two 1 GHz basebands, which consist of eight 128-MHz IFs, centered on 43 and 47 GHz, were employed. 
We averaged a total of 2 GHz bandwidth for a continuum imaging, hence the central frequency of 45 GHz (or 7 mm) was adopted in this paper. 

The primary calibrator 3C286 ($\sim$ 1.5 Jy) was observed for flux and bandpass calibrations at the same elevation. 
An accuracy of absolute flux scaling was better than 15\%, which was evaluated from a dispersion of $uv$-amplitude. 
In order to remove rapid atmospheric phase fluctuations at 45 GHz, we performed the fast-switching using a nearby ($\sim$ 1$\degr$.85) and bright ($\sim$ 1.5 Jy) quasar 1714-336.    
The switching cycle was set to 80 s (50 s for G353, 20 s for 1714-336, 10 s for slew). 
1714-336 was also used for pointing corrections every hour. 
Total on-source time was one hour. 

Data were analyzed using the Common Astronomy Software Applications (CASA) package. 
The VLA calibration pipeline was used for standard calibrations. 
The target was faint and no self-calibration was performed. 
Synthesized beam size (full width at half-maximum: FWHM) was 0$\arcsec$.13 $\times$ 0$\arcsec$.05, with a beam position angle of -0$\degr$.72 (east of north). 
Here, we adopted natural weighting for imaging, in order to maximize the sensitivity. 
This results in the final image sensitivity of 90 $\mu$Jy beam$^{-1}$ (1 $\sigma$). 
Although the typical astrometric accuracy for the J-VLA is 10\% of a synthesized beam in case of fast-switching, 
a positional error caused by thermal image noise reaches 6\% of the beam, because of a low signal-to-noise ratio (S/N $\sim$ 15 $\sigma$) in our case. 
We thus adopted a root sum square of both errors as the total astrometric error, i.e., $\sim$ 12 milliarcsecond (mas). 

\begin{deluxetable}{ll}
\tabletypesize{\scriptsize}
\tablecaption{Observational summaries\label{tbl-1}}
\tablewidth{0pt}
\tablehead{
\multicolumn2c{Phase-Tracking Center (J2000.0)} \\ 
\multicolumn1c{R.A.} & \multicolumn1c{Decl.}
}
\startdata
\multicolumn1c{17$^{\rm h}$26$^{\rm m}$01$^{\rm s}$.59} & \multicolumn1c{-34\degr15\arcmin14\arcsec.90}  \\ \hline
\multicolumn{2}{c}{J-VLA} \\ \hline
Observing date: & 2014 Feb 18  \\
Available antenna: & 24 -- 26 \\
Observing time: & 2.5 hr \\
On-source time: & 1 hr \\ 
Central frequencies: & 43 $/$ 47 GHz \\
Polarization: & dual \\
Bandwidth: & 1.0 GHz$/$IF\\ 
Primary calibrator: & 3C286 ($\sim$ 1.5 Jy) \\
Phase calibrator: & 1714-336($\sim$ 1.5 Jy)  \\
Synthesized beam: & 0\farcs13 $\times$ 0\farcs05 \\
Beam position angle: & -0\fdg72 \\
Image sensitivity ($1 \sigma$): & 90 $\mu$Jy beam$^{-1}$\\ 
Astrometric accuracy: & 12 mas \\ \hline
\multicolumn{2}{c}{ATCA} \\ \hline
Observing date: & 2013 Nov 1 and 3  \\
Available antenna: & 6 \\
Hour angle range: & $\pm$ 5 hr \\
On-source time: & 2.5 hr \\ 
Central frequency: & 6.668 GHz for maser\\ 
& 5.900 GHz for continuum \\ 
& 9.000 GHz for continuum \\
Polarization: & dual \\
Bandwidth: & 8.5 MHz for maser \\
& 2.0 GHz for continuum \\
Final spectral resolution: & 1.95 kHz (87.8 m s$^{-1}$) \\
Primary calibrator: & PKS1934-638 ($\sim$ 3.9 Jy) \\
Phase calibrator: & 1714-336 ($\sim$ 1.0 Jy)\\
Synthesized beam:\tablenotemark{a} & 2\farcs6 $\times$ 1\farcs4 for maser \\
& 2\farcs0 $\times$ 1\farcs0 for 6 GHz \\
& 1\farcs6 $\times$ 0\farcs8 for 9 GHz \\
Beam position angle: & 8\fdg6 for maser \\
& 5\fdg0 for 6 GHz \\
& 8\fdg4 for 9 GHz \\
Image sensitivity ($1 \sigma$): & 12 mJy beam$^{-1}$ for maser \\
& 64 $\mu$Jy beam$^{-1}$ for 6 GHz \\
& 46 $\mu$Jy beam$^{-1}$ for 9 GHz \\
Astrometric accuracy: & 400 mas \\ \hline
\enddata
\tablenotetext{a}{We adopted the uniform weighting for the maser and robust weighting for the continuum (see the main text). }
\end{deluxetable}

\subsection{ATCA observations} \label{sec:obs:atca}
Observations were held on 2013 November 1 and 3 in the 6A-configuration using all the six antennas. 
These observations were part of our imaging survey towards 10 candidates of southern DBSMs. 

The class II CH$_{3}$OH maser transition at 6.668519 GHz ($J_{K}$ = $5_{1}$ -- 6$_{0}$ A$^{+}$) was observed in the 1 MHz-0.5 k mode of the Compact Array broadband backend \citep{2011MNRAS.416..832W}. 
For the maser emission, a total bandwidth of 8.5 MHz in dual polarization centered on 6.668 GHz was obtained by concatenating 16 zoom bands. 
The spectral resolution was 0.488 kHz per channel and 4-channel smoothing was adopted for increased channel sensitivity. 
The final spectral resolution was 87.8 m s$^{-1}$ in the velocity domain. 
On the other hand, two 2 GHz wide data centered on 5.9 and 9.0 GHz were also obtained for continuum analysis. 

Several snapshot scans (10 - 15 minutes) were repeated within a hour angle range of $\pm$ 5 hr, acquiring better $uv$-coverage for each observing epoch. 
Total on-source time after flagging was 2.5 hr.   
We used the primary calibrator PKS1934-638 for flux and bandpass calibration. 
Phase and amplitude were calibrated by scanning 1714-336 ($\sim$ 1 Jy at 6.7 GHz) before and after every target scan. 
Array pointing was also checked using 1714-336 every hour. 

Basic reduction was performed by the MIRIAD package. 
The observed flux of PKS1934-638 ($\sim$ 3.9 Jy) was consistent with that estimated by interpolating 5.5 and 9 GHz fluxes in the ATCA Calibrator Database. 
We conclude that the accuracy of our flux scaling is better than 7\%. 
Self-calibration and imaging were conducted using the Astronomical Imaging Processing System (AIPS) package developed by the National Radio Astronomy Observatory (NRAO).

The maser data were self-calibrated by using the brightest maser channel. 
We adopted uniform weighting for maser mapping, achieving the maximum angular resolution. 
Synthesized beam size and position angles are 2$\arcsec$.6 $\times$ 1$\arcsec$.4 and 8$\degr$.6, respectively. 
The typical image noise level (1 $\sigma$) is $\sim$ 12 mJy beam$^{-1}$. 

On the other hand, we used robust weighting for the continuum in order to achieve a better image quality and a moderate angular resolution. 
The ROBUST parameter in AIPS task IMAGR was set to 0. 
This is just the medium weighting between the natural and uniform weighting. 
In order to remove strong contaminated flux from a nearby extended H II region, short baseline data were flagged out. 
We flagged all the data at a $uv$-distance shorter than 50 and 30 $k\lambda$ for 6 and 9 GHz band, respectively. 

Synthesized beam size and position angle are 2$\arcsec$.0 $\times$ 1$\arcsec$.0 and 5$\degr$.0 at 6 GHz, and 1$\arcsec$.6 $\times$ 0$\arcsec$.8 and 8$\degr$.4 at 9 GHz, respectively. 
These beam sizes are smaller than that in the uniform-weighted maser image, because of the baseline flagging mentioned above.  
The final image noise levels (1 $\sigma$) are $\sim$ 64 and 46 $\mu$Jy beam$^{-1}$ at 6 and 9 GHz, respectively.

The position of each maser spot and continuum source was determined by the elliptical Gaussian fitting. 
Although the absolute astrometric accuracy of ATCA is $\sim$ 400 mas, relative positional accuracies between detected maser spots are much better. 
Such internal accuracies, which are roughly estimated as the half of the beam size divided by S/N, reach a few mas in cases of bright maser spots.

\subsection{Archival IR data} \label{sec:obs:ir}
We retrieved the NASA/IPAC Infrared Science Archive website,\footnote{\url{http://irsa.ipac.caltech.edu/}}, searching for the nearest IR counterpart from the phase-tracking center. 
The retrieved catalogs are the Two Micron All Sky Survey (2MASS) Point Source Catalog \citep{2006AJ....131.1163S}, 
the {\it Spitzer} Galactic Legacy Infrared Mid-Plane Survey Extraordinaire \citep[GLIMPSE:][]{2003PASP..115..953B, 2009PASP..121..213C}, 
the mid-infrared {\it Midcourse Space eXperiment} ($MSX$) Point Source Catalog \citep{2001AJ....121.2819P, 2003yCat.5114....0E}, 
and the {\it Herschel} infrared Galactic Plane Survey \citep[Hi-GAL][]{2010A&A...518L.100M, 2016A&A...591A.149M} that consists of the {\it Herschel Space Observatory}  \citep{2010A&A...518L...1P} PACS \citep{2010A&A...518L...2P} and SPIRE \citep{2010A&A...518L...3G} images at wavelengths of 70, 160, 250, 350, and 500 $\mu$m. 
In addition to the IR dataset, we also found a submillimeter continuum data in the APEX Telescope Large Area Survey of the Galaxy (ATLASGAL) Compact Source Catalog
\footnote{\url{http://atlasgal.mpifr-bonn.mpg.de/cgi-bin/ATLASGAL_DATABASE.cgi}}\citep{2009A&A...504..415S, 2013A&A...549A..45C}. 

\section{Results} \label{sec:result}
\subsection{45 GHz Continuum}\label{sec:result:cont}
\begin{figure}[!htb]
\includegraphics[width=80mm]{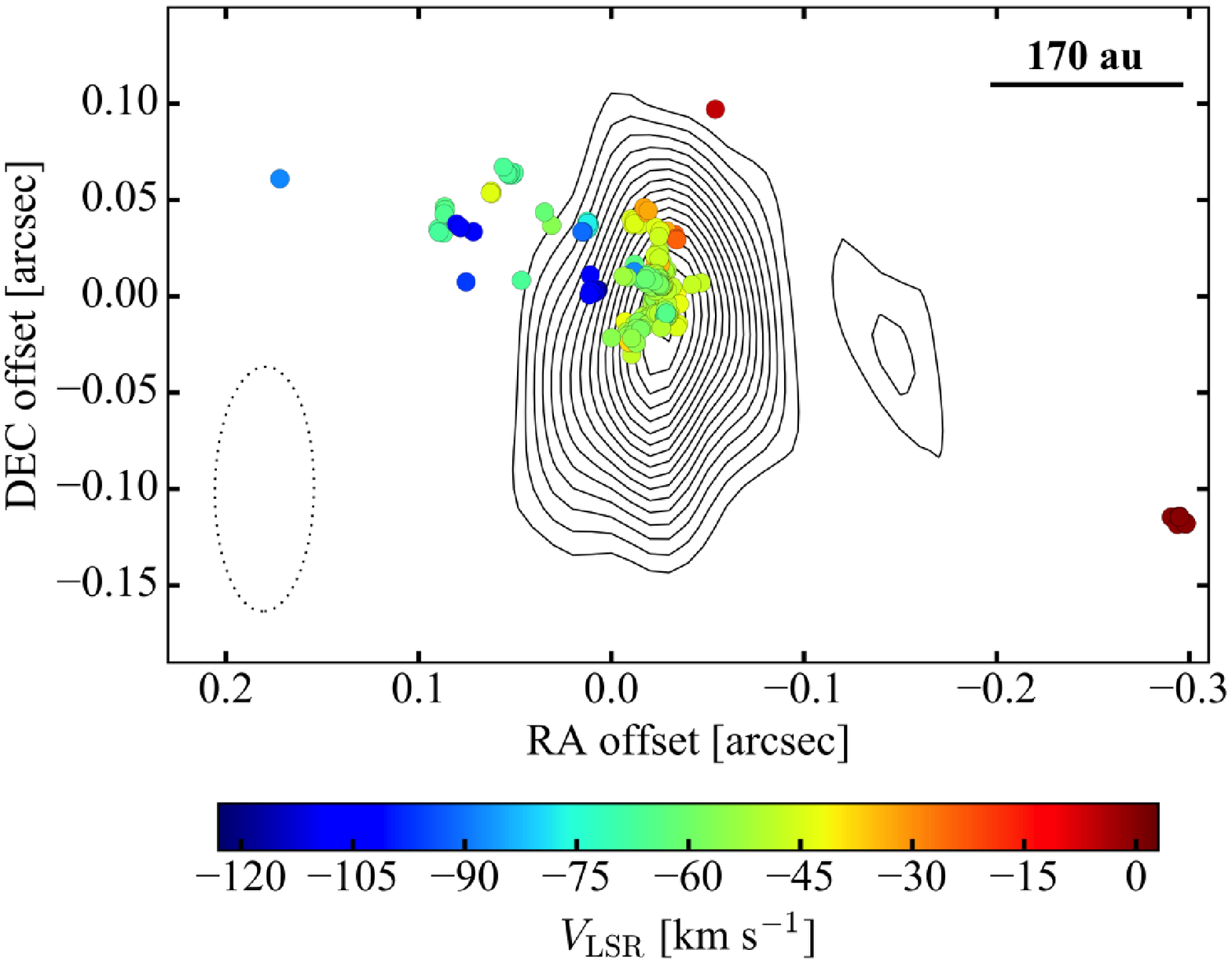}
\caption{The contours present J-VLA 45 GHz continuum image, which are from 19\% (3 $\sigma$) to 99\% with a step of 5\% of the image peak flux (1.39 mJy beam$^{-1}$). 
The filled circles show a VLBI map of the H$_2$O maser jet in \citet{2016PASJ...68...69M} with the color indicating the LOS velocity of each maser spot.  
The coordinate origin is the phase-tracking center, 17$^{\rm h}$26$^{\rm m}$01$^{\rm s}$.59, -34\degr15\arcmin14\arcsec.90 (J2000.0). 
The synthesized beam of J-VLA is shown in the lower left corner. \label{fig:1}}
\end{figure}

Figure \ref{fig:1} shows a 45 GHz continuum image taken by the J-VLA. 
In the same figure, we are also showing H$_{2}$O masers observed by VERA, which trace a protostellar jet \citep{2016PASJ...68...69M}. 
The $x$ and $y$ axes are the right ascension offset ($X$ =$\Delta\alpha$ $\times$ $cos\delta$) and declination offset ($Y$ =$\Delta\delta$), respectively. 
The coordinate origin was set to the phase-tracking center. 

A very compact continuum source was detected at the center of the bipolar maser jet. 
The source was resolved along the minor axis of the synthesized beam (in the east -west direction). 
Table \ref{tbl-2} shows the best-fit parameters that were derived from the elliptical Gaussian fitting. 
The total flux is 3.5 mJy and the size of the deconvolved Gaussian (FWHM: 0\arcsec.123 $\times$ 0\arcsec.073) indicates the averaged brightness temperature $T_{\rm b}$ of 235 K. 

Another weak component ($\sim$ 0.3 mJy beam$^{-1}$) was also detected at the west side. 
This can be a knot of the radio jet or binary companion, but we avoid further discussions due to the marginal S/N (3.5 $\sigma$). 

\begin{deluxetable*}{hhhcccccccc}
\tabletypesize{\scriptsize}
\tablecaption{The best-fit Gaussian parameters for the 45 GHz continuum \label{tbl-2}}
\tablewidth{0pt}
\tablehead{
\multicolumn{3}{h}{Image Size} & \multicolumn2c{Position Offset} & \multicolumn3c{Deconvolved Size (FWHM)} & \multicolumn3c{} \\
\nocolhead{$\theta_{\rm Maj}$} & \nocolhead{$\theta_{\rm Min}$} & \nocolhead{PA}&\colhead{$X$} & \colhead{$Y$} & \colhead{$\theta_{\rm Maj}$} & \colhead{$\theta_{\rm Min}$} & \colhead{PA}  & \colhead{$I_{\nu}$} & \colhead{$S_{\nu}$} & \colhead{$T_{\rm b}$}\\
\multicolumn{2}{h}{(mas)} & \nocolhead{(\degr)} & \multicolumn2c{(mas)} &\multicolumn2c{(mas)} & \colhead{(\degr)}&\colhead{(mJy beam$^{-1}$)} & \colhead{(mJy)} & \colhead{(K)} 
}
\decimals
\startdata
170 $\pm$ 2 & 101 $\pm$ 5 &165 $\pm$ 1 & -30 $\pm$ 1 & -16 $\pm$ 3 & 123 $\pm$ 4 & 73 $\pm$ 8  & 143 $\pm$ 5  & 1.3 $\pm$ 0.1 & 3.5 $\pm$ 0.1  & 235 $\pm$ 9 \\
\enddata
\tablecomments{Columns 1-2: positional offsets from the phase-tracking center. Columns 3-5: major axis, minor axis, and position angle (east of north). Columns 6-8: peak flux, total flux, and averaged brightness temperature. 
All the errors are formal fitting errors that do not include the astrometric error and the error in flux scaling. }
\end{deluxetable*}

\subsection{6/9 GHz Continuum}\label{sec:result:CX}
\begin{figure}[tb]
\includegraphics[width=80mm]{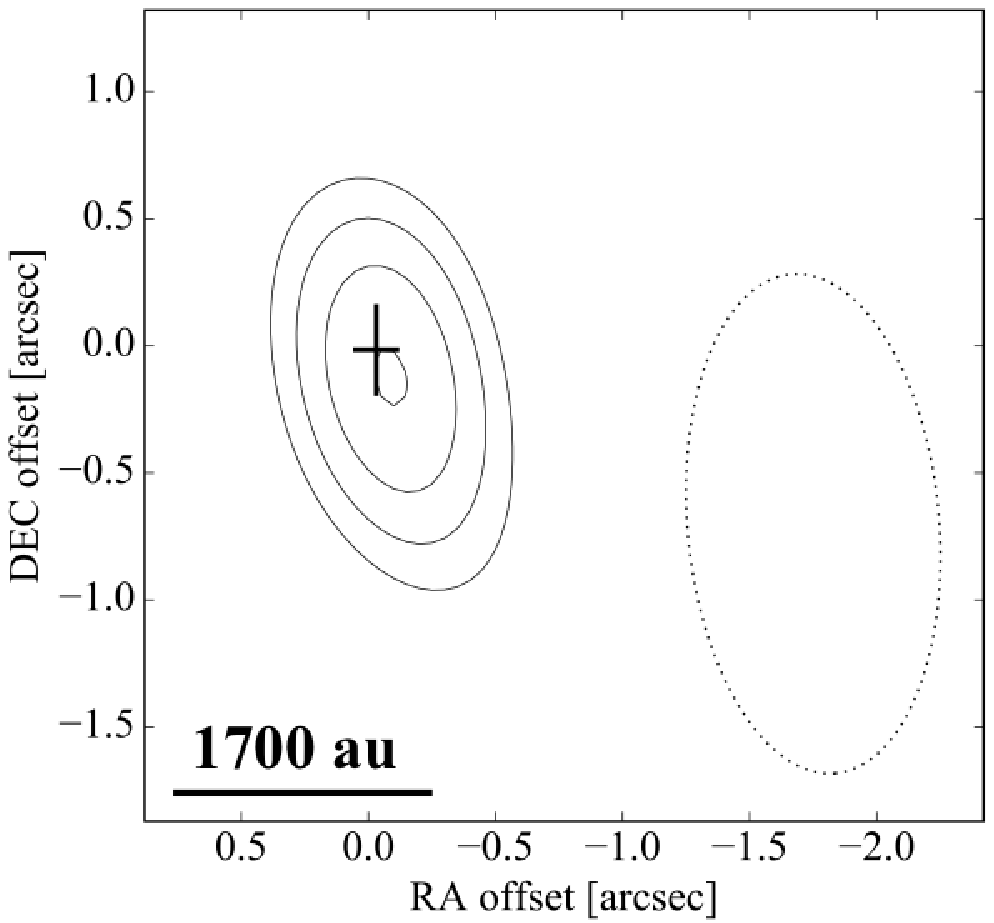}
\includegraphics[width=80mm]{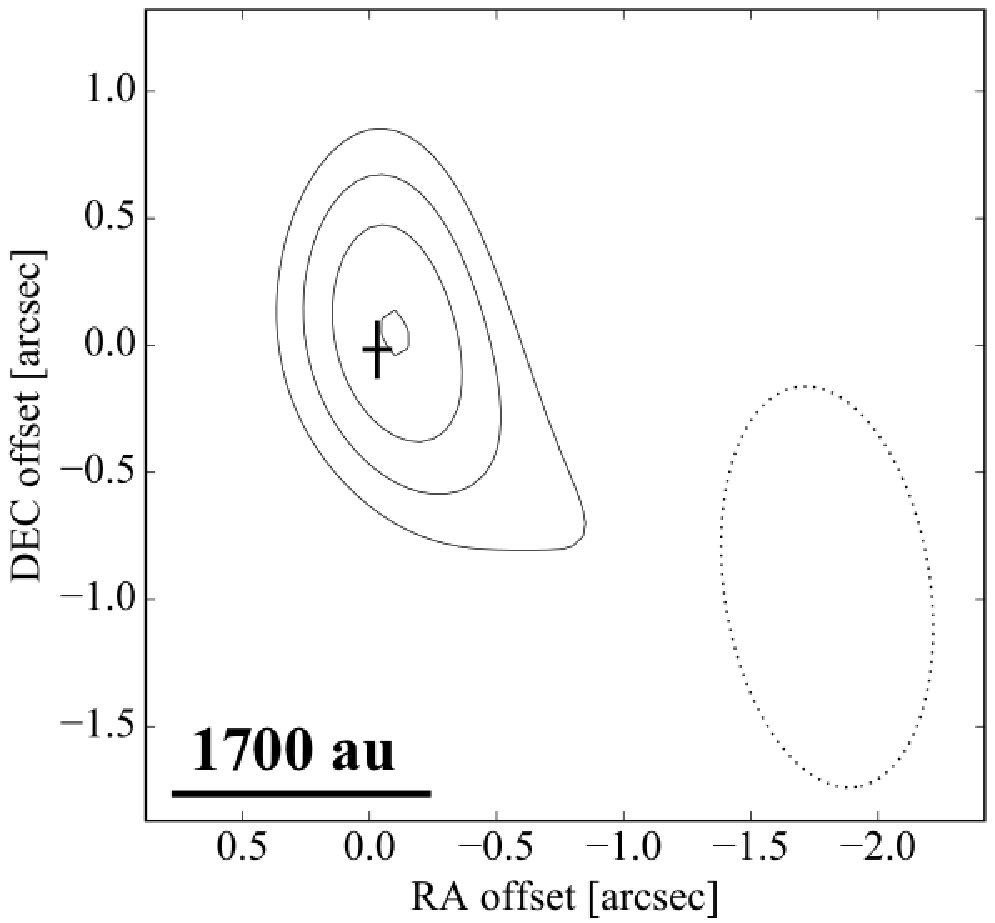}
\caption{Top: 6 GHz continuum image. Bottom: 9 GHz continuum image. 
The contour levels are 57\% (3 $\sigma$), 76\%, 94\%, and 99\% of the image peak flux (350 $\mu$Jy beam$^{-1}$) at 6 GHz, 
and 46\% (3 $\sigma$), 64\%, 81\%, and 99\% of the image peak flux (300 $\mu$Jy beam$^{-1}$) at 9 GHz. 
The black cross in each panel indicates the peak position at 45 GHz, where the cross size shows the relative positional errors (see Table \ref{tbl-3}). 
The dotted ellipse in each panel shows the relevant synthesized beam of ATCA. 
\label{fig:2}}
\end{figure}

We detected a faint centimeter continuum source (Figure \ref{fig:2}). 
Table \ref{tbl-3} presents properties of the 6/9 GHz continuum.  
The source positions are roughly consistent with that of the 45 GHz continuum emission within the astrometric error of 400 mas. 
Since the peak S/N is quite limited (5 $\sigma$ -- 7 $\sigma$), the relative positional errors are relatively large ($\sim$ 50 -- 200 mas). 
We note that the peak position of the best-fit gaussian at 9 GHz offsets from the image peak position (see Figure \ref{fig:2}), although the offset is still within the range of the fitting error ($\sim$ 100 mas). 
This is probably because of the slightly extended structure. 

 The source seems to completely be a point source at 6/9 GHz, while \citet{2013MNRAS.428..349M} have found that the radio jet is elongated $\sim$ 1$\arcsec$.0 along the north-south direction at 22 GHz. 
If one applies the size-frequency relation \citep{1996ASPC...93....3A}, where the major axis of a radio jet is inversely proportional to an observing frequency ($\theta_{\rm maj}\propto \nu^{-0.7}$ in the case of a conical jet), the source size is expected to be 2$\arcsec$.5 and 1$\arcsec$.9 at 6 and 9 GHz, respectively. 
Although this is comparable with the beam sizes in this work, it is clearly difficult to identify any partially resolved structure, because of the limited S/N. 

A tentative spectral index between 6 and 9 GHz is $\sim$ 0.6. 
This may be consistent with a typical index of a radio jet \citep[$\sim$ 0.6; ][]{1996ASPC...93....3A}, 
however, the error of the index is still significantly large ($\pm$ 1.3). 
The overall spectral energy distribution (SED) shape at centimeter wavelengths will be discussed in a latter section.

\begin{deluxetable*}{cccccccc}
\tabletypesize{\scriptsize}
\tablecaption{Properties of the 6/9 GHz continuum \label{tbl-3}}
\tablewidth{0pt}
\tablehead{
\multicolumn{1}{c}{Central} & \multicolumn4c{Position Offset\tablenotemark{a}} & \multicolumn3c{} \\
\colhead{Frequency} & \colhead{$X$} & \colhead{err} & \colhead{$Y$} & \colhead{err} & \colhead{$I_{\nu}$}  & \colhead{$S_{\nu}$}& \colhead{$uv$-limit\tablenotemark{b}} \\
 \colhead{(GHz)} &  \multicolumn2c{(mas)} & \multicolumn2c{(mas)} &  \colhead{($\mu$Jy beam$^{-1}$)} & \colhead{($\mu$Jy)} & \colhead{($k\lambda$)} 
}
\decimals
\startdata
5.9 & -99 & 91 & -154 & 181 & 350 $\pm$ 64 & 296 $\pm$ 100  & 50 \\
 9.0 & -200 & 58 & -104 & 114 & 320 $\pm$ 46 & 387 $\pm$ 91 & 30 \\
\enddata
\tablenotetext{a}{The peak position of the best-fit Gaussian with respect to the phase-tracking center. 
The errors do not include the absolute astrometric accuracy of ATCA.  }
\tablenotetext{b}{The minimum $uv$-distance for imaging (see the main text). }
\end{deluxetable*}

\begin{figure}[htb]
\epsscale{1.27}
\plotone{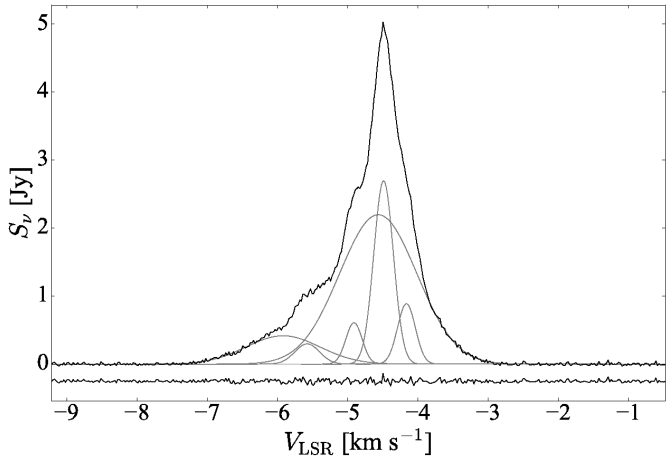}
\caption{ATCA spectrum of the 6.7 GHz CH$_{3}$OH maser. 
Data from all baselines were integrated. 
The thin curves show the six best-fit Gaussian components. 
The post-fit residual is also plotted with an offset of -0.25 Jy. 
\label{fig:3}}
\end{figure}

\subsection{6.7 GHz CH$_{3}$OH maser}\label{sec:result:maser}
The ATCA spectrum of the CH$_{3}$OH maser is shown in Figure \ref{fig:3}. 
The observed velocity range is roughly consistent with that reported in previous studies \citep[e.g., ][]{2008MNRAS.386.1521C, 2010MNRAS.404.1029C}, 
however, the spectral shape clearly differs from the double-peaked shape in the past. 
The peak flux is only 20\% compared to that in \citet{2008MNRAS.386.1521C}, where the relatively bright component ($\sim$ 25 Jy) was detected at -5 km s$^{-1}$. 

\begin{figure}[!tb]
\epsscale{1.15}
\plotone{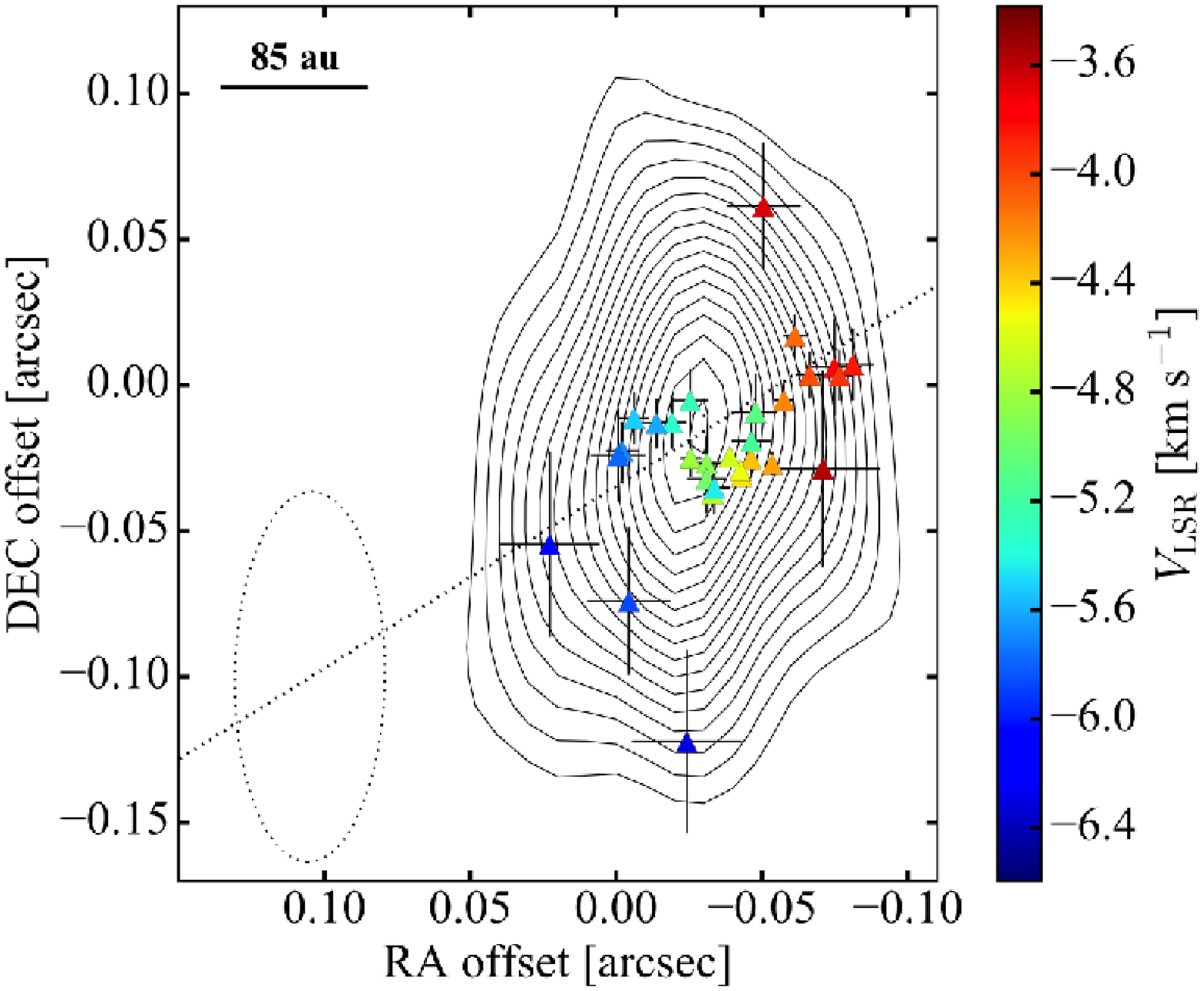}
\caption{Internal distribution of the 6.7 GHz CH$_{3}$OH masers. 
Filled triangle shows a position of each maser spot with a color indicating the LOS velocity. 
Only the bright maser spots (S/N $>$ 30 $\sigma$), which had better accuracy of an internal position, were shown here. 
The maser map is superposed on the 45 GHz continuum map, 
assuming that the peak position of the 6 GHz continuum is identical to that of the 45 GHz continuum. 
The contour levels are same as that in Figure \ref{fig:1}. 
The synthesized beam of J-VLA is shown in lower left corner. 
Black dotted line marks the axis of the position-velocity diagram. 
\label{fig:4}}
\end{figure}

We detected 43 maser spots with a 7 $\sigma$ detection limit (Table \ref{tbl-4}). 
The estimated positions of all the maser spots are roughly consistent with the position of the 45 GHz continuum emission within the astrometric error of 400 mas, 
although their averaged position is slightly offset from the phase-tracking center, i.e., -104 and -160 mas in $X$ and $Y$, respectively. 
This offset is almost the same as that of the 6 GHz band continuum. 
Figure \ref{fig:4} presents the internal distribution of 30 bright CH$_{3}$OH maser spots (S/N $>$ 30), where the coordinated origin was slightly offset from the original position (see below).  
This S/N corresponds to the relative positional accuracy of $\sim$ 25 and 45 mas in the $X$ and $Y$ directions, respectively. 
These masers have accurate relative positions compared to the scale of the maser distribution, and hereafter we use only these spots for discussing detailed internal structure. 

\begin{deluxetable*}{crrrrrr}
\tablecaption{Detected maser spots \label{tbl-4}}
\tablehead{\colhead{$V_{\rm LSR}$} & \multicolumn1c{$X$} & \multicolumn1c{$Y$} & \colhead{$R$} & \multicolumn1c{$I_{\nu}$} & \multicolumn1c{$S_{\nu}$} & \colhead{S/N} \\ 
\colhead{(km s$^{-1}$)} & \multicolumn1c{(mas)}&\multicolumn1c{(mas)}& \colhead{(mas)} & \multicolumn1c{(Jy beam$^{-1}$)} & \multicolumn1c{(Jy)} & \colhead{} } 
\startdata
\multicolumn{7}{c}{Bright Spots (S/N $\geq$ 30)} \\ \hline
-3.584  & -35  $\pm$ 20  & -6  $\pm$ 34  & 36  & 0.427  $\pm$ 0.014  & 0.416  $\pm$ 0.023  & 31  \\
-3.672  & -15  $\pm$ 13  & 84  $\pm$ 21  & 85  & 0.594  $\pm$ 0.012  & 0.578  $\pm$ 0.021  & 48  \\
-3.759  & -39  $\pm$ 9  & 28  $\pm$ 17  & 48  & 0.830  $\pm$ 0.013  & 0.841  $\pm$ 0.022  & 65  \\
-3.847  & -45  $\pm$ 7  & 29  $\pm$ 12  & 54  & 1.095  $\pm$ 0.013  & 1.097  $\pm$ 0.022  & 86  \\
-3.935  & -41  $\pm$ 5  & 26  $\pm$ 9  & 48  & 1.422  $\pm$ 0.011  & 1.446  $\pm$ 0.020  & 125  \\
-4.023  & -31  $\pm$ 5  & 26  $\pm$ 8  & 40  & 1.705  $\pm$ 0.013  & 1.711  $\pm$ 0.022  & 135  \\
-4.111  & -26  $\pm$ 4  & 39  $\pm$ 7  & 47  & 1.905  $\pm$ 0.013  & 1.922  $\pm$ 0.022  & 151  \\
-4.198  & -22  $\pm$ 3  & 17  $\pm$ 6  & 28  & 2.379  $\pm$ 0.012  & 2.381  $\pm$ 0.021  & 193  \\
-4.286  & -18  $\pm$ 2  & -5  $\pm$ 4  & 18  & 3.729  $\pm$ 0.014  & 3.707  $\pm$ 0.024  & 274  \\
-4.374  & -10  $\pm$ 1  & -3  $\pm$ 3  & 11  & 6.309  $\pm$ 0.015  & 6.289  $\pm$ 0.026  & 426  \\
-4.462  & -7  $\pm$ 1  & -9  $\pm$ 2  & 12  & 8.617  $\pm$ 0.017  & 8.579  $\pm$ 0.029  & 510  \\
-4.550  & -7  $\pm$ 1  & -7  $\pm$ 2  & 10  & 8.324  $\pm$ 0.017  & 8.314  $\pm$ 0.029  & 493  \\
-4.637  & -3  $\pm$ 1  & -2  $\pm$ 3  & 4  & 6.128  $\pm$ 0.015  & 6.138  $\pm$ 0.026  & 417  \\
-4.725  & 3  $\pm$ 2  & -15  $\pm$ 4  & 15  & 3.719  $\pm$ 0.013  & 3.751  $\pm$ 0.023  & 284  \\
-4.813  & 10  $\pm$ 4  & -3  $\pm$ 6  & 11  & 2.187  $\pm$ 0.013  & 2.198  $\pm$ 0.023  & 167  \\
-4.901  & 5  $\pm$ 5  & -5  $\pm$ 9  & 7  & 1.473  $\pm$ 0.013  & 1.462  $\pm$ 0.022  & 117  \\
-4.989  & 5  $\pm$ 7  & -10  $\pm$ 12  & 11  & 1.157  $\pm$ 0.012  & 1.183  $\pm$ 0.022  & 93  \\
-5.076  & -12  $\pm$ 8  & 13  $\pm$ 13  & 18  & 1.059  $\pm$ 0.013  & 1.078  $\pm$ 0.023  & 82  \\
-5.164  & -11  $\pm$ 7  & 3  $\pm$ 12  & 11  & 1.108  $\pm$ 0.012  & 1.142  $\pm$ 0.022  & 91  \\
-5.252  & 10  $\pm$ 6  & 17  $\pm$ 11  & 20  & 1.279  $\pm$ 0.013  & 1.273  $\pm$ 0.023  & 98  \\
-5.340  & 17  $\pm$ 5  & 9  $\pm$ 9  & 19  & 1.420  $\pm$ 0.012  & 1.406  $\pm$ 0.020  & 121  \\
-5.428  & 2  $\pm$ 5  & -13  $\pm$ 9  & 13  & 1.495  $\pm$ 0.013  & 1.477  $\pm$ 0.022  & 119  \\
-5.515  & 30  $\pm$ 5  & 11  $\pm$ 9  & 32  & 1.541  $\pm$ 0.013  & 1.599  $\pm$ 0.023  & 121  \\
-5.603  & 22  $\pm$ 5  & 9  $\pm$ 9  & 24  & 1.539  $\pm$ 0.012  & 1.562  $\pm$ 0.021  & 126  \\
-5.691  & 34  $\pm$ 6  & 0  $\pm$ 10  & 34  & 1.344  $\pm$ 0.012  & 1.350  $\pm$ 0.022  & 108  \\
-5.779  & 35  $\pm$ 9  & -2  $\pm$ 16  & 35  & 0.903  $\pm$ 0.013  & 0.933  $\pm$ 0.024  & 68  \\
-5.867  & 31  $\pm$ 14  & -52  $\pm$ 25  & 61  & 0.523  $\pm$ 0.012  & 0.554  $\pm$ 0.021  & 44  \\
-6.218  & 59  $\pm$ 17  & -32  $\pm$ 32  & 67  & 0.420  $\pm$ 0.012  & 0.421  $\pm$ 0.021  & 35  \\
-6.306  & 12  $\pm$ 19  & -100  $\pm$ 32  & 101  & 0.445  $\pm$ 0.013  & 0.486  $\pm$ 0.023  & 35  \\
-6.394  & 48  $\pm$ 19  & -50  $\pm$ 33  & 69  & 0.422  $\pm$ 0.012  & 0.470  $\pm$ 0.023  & 34  \\ \hline
\multicolumn{7}{c}{Faint Spots (S/N $<$ 30)} \\ \hline
-3.232  & -88  $\pm$ 59  & -23  $\pm$ 112  & 91  & 0.124  $\pm$ 0.012  & 0.131  $\pm$ 0.022  & 10  \\
-3.320  & -72  $\pm$ 39  & 77  $\pm$ 65  & 105  & 0.198  $\pm$ 0.012  & 0.190  $\pm$ 0.021  & 16  \\
-3.408  & -29  $\pm$ 30  & 112  $\pm$ 50  & 116  & 0.270  $\pm$ 0.013  & 0.252  $\pm$ 0.022  & 20  \\
-3.496  & -148  $\pm$ 26  & -8  $\pm$ 44  & 148  & 0.305  $\pm$ 0.013  & 0.304  $\pm$ 0.022  & 24  \\
-5.955  & -20  $\pm$ 23  & -36  $\pm$ 38  & 42  & 0.331  $\pm$ 0.012  & 0.353  $\pm$ 0.022  & 28  \\
-6.042  & 36  $\pm$ 26  & 23  $\pm$ 45  & 42  & 0.291  $\pm$ 0.012  & 0.303  $\pm$ 0.021  & 24  \\
-6.130  & 23  $\pm$ 24  & -72  $\pm$ 40  & 75  & 0.323  $\pm$ 0.012  & 0.323  $\pm$ 0.022  & 26  \\
-6.481  & -21  $\pm$ 26  & -50  $\pm$ 40  & 54  & 0.317  $\pm$ 0.012  & 0.337  $\pm$ 0.022  & 26  \\
-6.569  & 19  $\pm$ 38  & -33  $\pm$ 65  & 38  & 0.219  $\pm$ 0.013  & 0.221  $\pm$ 0.022  & 17  \\
-6.657  & 103  $\pm$ 50  & -41  $\pm$ 96  & 111  & 0.145  $\pm$ 0.012  & 0.150  $\pm$ 0.021  & 12  \\
-6.745  & 40  $\pm$ 51  & -179  $\pm$ 103  & 183  & 0.130  $\pm$ 0.012  & 0.124  $\pm$ 0.020  & 11  \\
-6.833  & 91  $\pm$ 59  & 121  $\pm$ 105  & 151  & 0.111  $\pm$ 0.012  & 0.085  $\pm$ 0.018  & 9  \\
-6.920  & -49  $\pm$ 119  & -56  $\pm$ 177  & 74  & 0.086  $\pm$ 0.011  & 0.115  $\pm$ 0.024  & 8  \\
\enddata
\tablecomments{Column 1: LOS velocities with respect to the local standard of rest (LSR). Columns 2-3: positional offsets from the averaged position, 
that is, -104 and -160 mas in $X$ and $Y$ with respect to the phase-tracking center, respectively. 
Column 4: projected distances from the averaged position. Columns 5-6: Peak and integrated fluxes.  Column 7: S/Ns. 
All the errors are formal fitting errors that do not include the astrometric error and the error in flux scaling. }
\end{deluxetable*}

It is difficult to directly compare the 45 GHz continuum position and maser distribution, due to the significant astrometric error in ATCA data. 
Therefore, we superposed the maser distribution onto the continuum image in Figure \ref{fig:4}, 
assuming that the peak position of the 6 GHz continuum is identical to that of the 45 GHz continuum. 
In this assumption, the spatial spread of the maser cluster coincides with that of 45 GHz continuum emission, even including fainter (S/N $<$ 30) masers. 
This could suggest that both emissions stem from the same circumstellar structure. 

\begin{figure}[!htb]
\epsscale{1.2}
\plotone{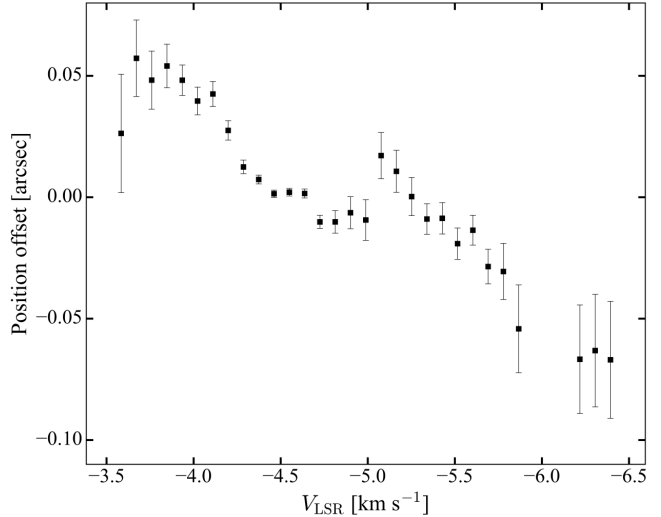}
\caption{Position-velocity diagram of the CH$_{3}$OH masers. 
The $x$ axis shows a LOS velocity. 
The $y$ axis shows a positional offset from the averaged position along the the position angle of 122$\degr$ that was estimated by the simple linear fitting. 
\label{fig:5}}
\end{figure}

It is noteworthy that the masers show a systematic velocity gradient along the inverse-"S" shaped alignment. 
Figure \ref{fig:5} shows the position-velocity diagram. 
Here, the vertical axis shows a positional offset from the averaged position, along the position angle of 122$\degr$ that was estimated by the simple linear fitting toward the maser distribution. 
The velocity gradient is 2.0 $\times$ 10$^{-2}$ km s$^{-1}$ mas$^{-1}$  ($\pm$ 1.5 km s$^{-1}$ over 150 mas) centered on the systemic velocity of -5 km s$^{-1}$. 
The direction of the velocity gradient (southeast-northwest) is roughly perpendicular to the position angle of the H$_{2}$O maser jet (northeast-southewest). 
This probably suggests a different kinematic origin for the two masers.

\subsection{Bolometric luminosity}\label{sec:result:bol}
\begin{figure}[htb]
\epsscale{1.25}
\plotone{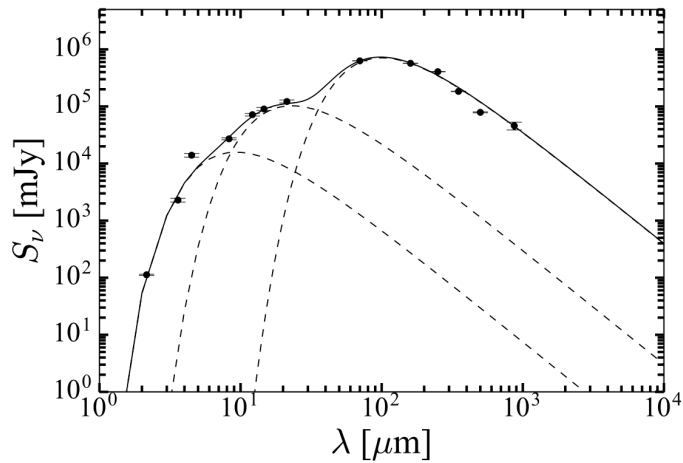}
\caption{Best-fit result of the three-component blackbody fitting with the extinction correction ($A_{\rm v}$ = 20 mag). 
The dashed and solid curves show each blackbody component ($T$ = 51, 224, and 550 K) and total flux, respectively. 
\label{fig:6}}
\end{figure}

Table \ref{tbl-5} presents summaries of the IR archival data. 
Unfortunately, two GLIMPSE data at 5.8 and 8.0 $\mu$m were saturated and out of use. 
Figure \ref{fig:6} shows the SED at IR wavelengths. 
Here, foreground extinction was corrected, based on a visual extinction ($A_{\rm V}$) of $\sim$ 20 mag in this region, which was estimated by \citet{2010AA...515A..55R}. 
A dust opacity model in \citet{1994A&A...291..943O} without any coagulation was adopted, assuming a less dense extended cloud as a foreground opacity source.  

An important feature is a relatively flat SED shape. 
The mid-IR fluxes at 10-20 $\mu$m reach $\sim$ 10\% of the peak flux around 100 $\mu$m. 
Such a flat SED is theoretically expected for a nearly face-on object \citep{2011ApJ...733...55Z} and consistent with the previously measured inclination angle. 

In order to estimate a bolometric luminosity ($L_{\rm bol}$), the SED shape was approximated by a simple three-component blackbody model. 
We performed this blackbody fitting using both of the extinction-corrected data and the original data for comparison. 
The fitting results are summarized in Table \ref{tbl-6}. 
The estimated $L_{\rm bol}$ is $\sim$ 5$\times$10$^{3}$ $L_{\sun}$ and only weakly depends on the extinction within the adopted $A_{\rm V}$ range (0 --20 mag). 
This corresponds to a B1-type zero-age main sequence star (ZAMS) \citep[e.g.,][]{1973AJ.....78..929P}, implying a stellar mass of $\sim$ 10 $M_{\sun}$ \citep[e.g.,][]{2009ApJ...691..823H}. 

This is the first estimation of a host YSO mass in G353 and we adopt 10 $M_{\sun}$ in the following discussions. 
However, it should be noted that this is still a rough estimation. 
A fine-tuned SED modeling that includes a stellar evolutionary model is required to conclude exact stellar parameters.

\begin{deluxetable*}{llrcrrrr}
\tabletypesize{\scriptsize}
\tablecaption{IR archival Data Set \label{tbl-5}}
\tablewidth{0pt}
\tablehead{
 \multicolumn1l{Catalog} & \multicolumn1l{Source ID} & \multicolumn1c{$\lambda$ } &\multicolumn1l{Separation$\tablenotemark{a}$} & \multicolumn1c{$S_{\rm \nu}$} & \multicolumn1c{err} & \multicolumn1c{$S^{\prime}_{\rm \nu}$} & \multicolumn1c{err$^{\prime}$} \\  \cline{5-8}
 \multicolumn1l{}  & \colhead{}& \multicolumn1c{($\mu$m)}  & \colhead{(arcsec)}  & \multicolumn4c{(Jy)} 
}
\decimals
\startdata
2MASS & 17260170-3415154 & 2.2 & 1.5 & 0.0062 & 0.0001 & 0.091 & 0.002 \\ \hline
GLIMPSE & G353.2732+00.6409 & 3.6 & 0.7 & 0.64 & 0.05 & 2.04 & 0.16 \\ 
          & & 4.5 & & 5.3 & 0.4 & 12.9 & 1.0 \\
           & & 5.8 & & \multicolumn4c{Saturated}  \\
           & & 8.0 & & \multicolumn4c{Saturated}  \\ \hline
MSX & G353.2732+00.6411 &  8.3 & 0.5 &13.7 & 0.6 & 25.7 & 1.0 \\
          &&  12.1 & & 28.7 & 1.4  & 64.7 & 3.2 \\
          &&  14.7 & & 51.7 & 3.2   & 85.3 & 5.2 \\ 
          &&  21.3 & & 67.2 & 4.0  & 111.9 & 6.7 \\ \hline
 Hi-GAL & HIGALPB353.2731+0.6416 &  70 & 1.6 & 590.3 & 0.6 & 621.2 & 0.7 \\
& HIGALPR353.2733+0.6418 &  160 & 2.6 &560.2 & 1.3  & 567.5 & 1.4 \\
& HIGALPS353.2731+0.6418  &  250 & 2.4 &402.8 & 1.6 & 405.3 & 1.6 \\
& HIGALPM353.2734+0.6421 &   350 & 3.8 &182.7 & 1.5  & 183.2 & 1.5\\
& HIGALPL353.2737+0.6429 &  500 & 6.7 & 78.7 & 1.5  & 78.8 & 1.5 \\ \hline
ATLASGAL  & AGAL353.272+00.641 &  870 & 3.0 & 45.9 & 0.7 & 45.9 & 0.7  \\
\enddata
\tablenotetext{a}{Angular separation from the phase-tracking center. }
\tablecomments{The last two columns show extinction-corrected fluxes and errors, respectively. } 
\end{deluxetable*}

\begin{deluxetable}{cccc}
\tabletypesize{\scriptsize}
\tablecaption{Approximated bolometric luminosity\label{tbl-6}}
\tablewidth{0pt}
\tablehead{
$A_{\rm v}$ & $N_{\rm H_{2}}$ & $L_{\rm bol}$  & Comments \\
 \colhead{(mag)} & \colhead{(10$^{22}$ cm$^{-2}$)}&  \colhead{(10$^{3}$ $L_{\sun}$)} & \colhead{} 
}
\decimals
\startdata
0   & 0.0  & 4.5 & No extinction correction\\
20 &  2.0 & 5.5  & $A_{\rm v}$ from  \cite{2010AA...515A..55R} \\
\enddata
\tablecomments{The first two columns show a foreground visual extinction and corresponding H$_{2}$ column density. } 
\end{deluxetable}

\section{Discussion}\label{sec:discus}
\subsection{Origin of the continuum emission}\label{sec:discus:cont} 
\begin{figure}[!htb]
\epsscale{1.22}
\plotone{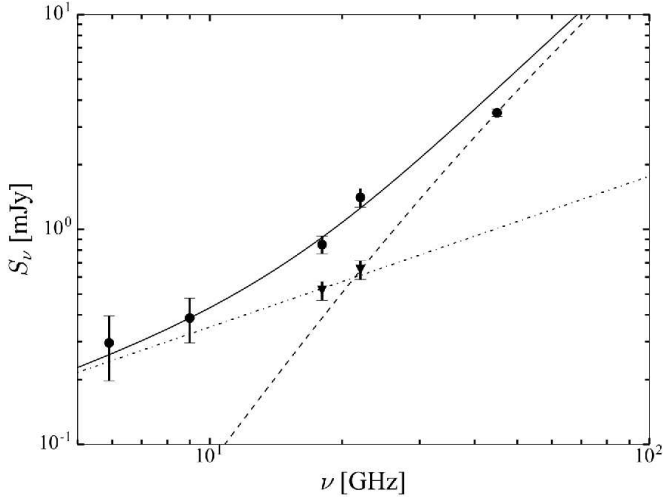}
\caption{Tentative SED of G353 between 6 and 45 GHz. The filled circles present total fluxes, including 18 and 22 GHz data from \citet{2013MNRAS.428..349M}. 
The two downward triangles also present the peak fluxes at 18 and 22 GHz that were treated as the upper limit fluxes for the compact continuum source. 
The solid black curve shows the best-fit "free-free + dust" model for $\beta$ = 1.00 (see the main text); on the other hand, the dash-dotted and dashed curve indicate each of the free-free and dust components in the best-fit model, respectively.
\label{fig:7}}
\end{figure}

Figure \ref{fig:7} shows a tentative SED at centimeter wavelengths. The SED contains both of the peak and total fluxes at 22 GHz continuum reported in \citet{2013MNRAS.428..349M}. 
The former gives an upper limit of the compact continuum source, while the latter, and also 6/9 GHz fluxes, contain significant contributions from the extended radio jet because of the lower angular resolution  (0$\arcsec$.5 -- 2$\arcsec$.0). 
The upper limit flux at 22 GHz suggests that the lower limit spectral index $\alpha$ of the compact continuum source is 2.4 $\pm$ 0.2 between 22 and 45 GHz. 

This could exclude any possibility of a very compact H II region \citep[$\alpha$ = -0.1 -- 2:][]{2010MNRAS.405.1560M} or radio jet \citep[$\alpha$ $\sim$ 0.7:][]{1996ASPC...93....3A}. 
A fully optically thick hyper-compact H II region ($\alpha$ = 2), which is occasionally associated with the class II CH $_{3}$OH maser \citep[e.g., ][]{2011ApJ...739L...9S}, may be still possible if there are significant errors in the absolute flux scaling. 
However, the observed flux of 3.5 mJy at 45 GHz corresponds to a source size of $\sim$ 18 mas in this case, considering the electron temperature of 10${^4}$ K. 
This is clearly inconsistent with the fact that the source was slightly resolved in the J-VLA image.

Alternatively, such a steep index is naturally explained by a graybody emission from dust ($\alpha$ = 2 -- 4). 
The association between the continuum and 6.7 GHz CH$_{3}$OH maser, which is excited by warm dust emission \citep{2005MNRAS.360..533C}, is also consistent in this case. 
We thus propose that the continuum traces warm dust in the innermost circumstellar system. 

No significant beam dilution is expected, since the source structure was clearly resolved. 
The observed brightness temperature of 235 K implies that an optical depth of $\tau$ is at least higher than 0.17, considering the dust sublimation temperature ($\sim$ 1500 K). 
This gives a valid lower limit gas mass of 0.2 $M_{\sun}$, assuming typical dust parameters in an accretion disk, i.e., 
dust mass opacity at 350 GHz $k_{\rm 350 GHz}$ = 1.75 cm$^{2}$ g$^{-1}$, opacity index $\beta$ = 1.0, and a gas-to-dust ratio of 100 \citep[e.g.,][]{1994A&A...291..943O, 2007ApJ...659..479J}. 
The dust mass opacity at a frequency $\nu$ is expressed as $k_{\nu}$ = 1.75 ($\nu$$/$$\rm 350\:GHz$)$^{\beta}$ cm$^{2}$ g$^{-1}$. 

On the other hand, a partially optically thick condition ($\tau$ $\sim$ 1) gives a more moderate dust temperature ($T_{\rm dust} \sim$ 375 K). 
Such an optically thick condition may be consistent with the "disk-masking" scenario for the origin of DBSMs proposed by \citet{2013MNRAS.428..349M}, 
where an optically thick face-on disk obscures most of the red-shifted H$_{2}$O maser spots. 
We tried a simple model fitting toward the overall SED shape, in order to test such a possibility. 

We considered a two-component SED model combining a radio jet and dust emission, where total flux was defined by, 
\begin{eqnarray*}
S_{\nu}  & = & S_{\rm jet} + S_{\rm dust} \\
               & = & S_{\rm 22}\left(\frac{\nu}{\rm 22 GHz}\right )^{\alpha_{\rm jet}} + \Omega_{\rm dust}\left(1 - \exp(-\tau_{\rm dust})\right)B_{\nu}(T_{\rm dust}). 
\end{eqnarray*}
Here, $S_{\rm 22}$ and $\alpha_{\rm jet}$ are the contributed flux of the radio jet at 22 GHz and the spectral index of the jet, respectively. 
$\Omega_{\rm dust}$ is a solid angle of the compact dust emission and we adopted the observed source size in Table \ref{tbl-2}. 
$B_{\nu}(T_{\rm dust})$ is a Planck function. 
The optical depth of the dust emission $\tau_{\rm dust}$ is expressed as, 
\begin{eqnarray*}
\tau_{\rm dust}  & = & \tau_{\rm 45}\left(\frac{\nu}{\rm 45 GHz}\right )^{\beta}, 
\end{eqnarray*}
where $\tau_{\rm 45}$ indicates the optical depth of the dust emission at 45 GHz. 
We assumed $S_{\rm dust}$ = 3.5 mJy (i.e., $T_{\rm b}$ = 235 K) at 45 GHz. 
This is because any extended flux at 45 GHz seems to be completely resolved out. 
$\beta$ could not be determined uniquely, because of an optically thick condition. 
Instead, the observed SED was fitted, adopting several fixed $\beta$ values ($\beta$ = 1.0, 0.75, and 0.5). 
The remaining three parameters ($\alpha_{\rm jet}$,  $S_{22}$ and $\tau_{\rm 45}$) were determined under these assumptions. 

The best-fit parameters were summarized in Table \ref{tbl-7} and each of $S_{\nu}$, $S_{\rm jet}$ and $S_{\rm dust}$ in the case of $\beta$ = 1.0 were each plotted in Figure \ref{fig:7}. 
The obtained $\alpha_{\rm jet}$ and $S_{22}$ were almost identical in all the assumed $\beta$. 
In particular, the best-fit $\alpha_{\rm jet}$ of 0.7 is well consistent with the typical value of the radio jet \citep{1996ASPC...93....3A}. 
The dust emission is, on the other hand, expected to be optically thick in all cases. 
Table \ref{tbl-7} also contains the resultant $T_{\rm dust}$ and total gas masses $M_{\rm gas}$ for each $\beta$. 
The latter were calculated assuming a gas-to-dust ratio of 100 and dust mass opacity in \citet{1994A&A...291..943O} mentioned above. 

The estimated $M_{\rm gas}$ is 0.6 $M_{\sun}$ even in the case of $\beta$ = 0.5 and up to a few $M_{\sun}$, 
unless the gas-to-dust ratio and/or dust opacity are significantly lower than the nominal interstellar conditions, as in the cases of low-mass accretion disks \citep[e.g.,][and references therein]{2017ApJ...838..151T}. 
If $\beta$ is larger than 0.75, $M_{\rm gas}$ exceeds 15\% of the host YSO mass ($\sim$ 10 $M_{\sun}$). 
Such a High-Mass system could be self-gravitating and dynamically unstable. 
This may explain the episodic nature of the H$_{2}$O maser jet \citep{2016PASJ...68...69M}, 
by causing a short-time ($\sim$ 1 year) variation of the accretion rate \citep[e.g.,][and references therein]{2014ApJ...796L..17M}, as suggested in recent NIR observations toward several HMYSOs \citep[e.g.,][]{2016ApJ...833...24K}. 
Otherwise, the results may simply imply smaller $\beta$ in the system. 

Our SED analysis has shown that the observed SED of G353 could be explained by a combination of an extended radio jet and compact optically thick dust source; 
however, we emphasize that the expected physical parameters are still tentative and depend strongly on the dust parameters. 
In addition, we cannot exclude any partial contamination from an unresolved free-free emission around the host object. 
We require a follow-up SED measurements at a comparable or higher angular resolution to confirm the exact nature of the circumstellar environment in G353. 
Lower-frequency observations at optically thin frequencies are particularly important to determine precise dust parameters and column density.

\begin{deluxetable}{cc|ccc|cc}
\tabletypesize{\scriptsize}
\tablecaption{The best-fit SED parameters \label{tbl-7}}
\tablewidth{0pt}
\tablehead{
\multicolumn{2}{c|}{Fixed} & \multicolumn{3}{c}{Free} & \multicolumn{2}{|c}{Results} \\  \cline{1-7} 
\multicolumn{1}{c}{$\beta$} & \multicolumn{1}{c|}{$T_{\rm b}$\tablenotemark{a}} &\colhead{$\alpha_{\rm jet}$} & \colhead{$S_{22}$} & \colhead{$\tau_{45}$} &  \multicolumn{1}{|c}{$T_{\rm dust}$}&  \colhead{$M_{\rm gas}$\tablenotemark{b}}  \\
\colhead{}  & \multicolumn{1}{c|}{(K)}& \colhead{} & \colhead{(mJy)} &  \colhead{}  & \multicolumn{1}{|c}{(K)} &  \colhead{($M_{\sun}$)}  
}
\decimals
\startdata
1.00 & 235&  0.7 & 0.6 &  2.5 &  258  & 3.9 \\
0.75 & 235 & 0.7 & 0.6 &  1.8 &  284  & 1.7 \\
0.50 & 235 & 0.7 & 0.6 &  1.0 &  374  & 0.6 \\
\enddata
\tablecomments{}
\tablenotetext{a}{The observed brightness temperature at 45 GHz. }
\tablenotetext{b}{A gas-to-dust ratio of 100 was assumed. }
\end{deluxetable}

\subsection{Origin of the velocity gradient}\label{sec:discus:maser}
Although similar systematic velocity gradients are found in some portion of class II CH$_{3}$OH maser sites \citep[e.g., ][]{1998ApJ...508..275N, 1998MNRAS.301..640W, 2009A&A...502..155B, 2014PASJ...66...31F}, 
one should be careful that our beam size is still significantly larger than the maser distribution. 
This could artificially create such a velocity gradient in some cases. 
For example, if two maser clumps with slightly different velocity centroids are closely located within a beam size, systematic drift of the apparent velocity centroid may be observed. 
However, this is probably not the case, since the interferometric spectrum in Figure \ref{fig:3} consists of at least six Gaussian components. 
This fact suggests that the masers have intrinsic kinematic and spatial structures.

The linear position-velocity diagram in Figure \ref{fig:5} may be explained by an edge-on Keplerian rotating ring \citep[e.g.,][]{2007ApJ...663..857U}, 
despite the nearly face-on geometry of the jet as shown in \citet{2016PASJ...68...69M}. 
Such an apparently misaligned geometry might be possible, if the jet significantly bends within 100 au from the launching point. 
However, the observed velocity gradient (3 km s$^{-1}$ over 250 au) corresponds to a dynamical mass of only 0.3 $M_{\sun}$ in this case. 
This is too small and clearly inconsistent with the association of the class II CH$_{3}$OH maser itself. 
In addition, a simple spherical infall model is also unsuitable, since LOS velocities coincide with the systemic velocity of $\sim$ -5 km s$^{-1}$ near the center of the distribution. 

\begin{figure}[tb]
\epsscale{1.21}
\plotone{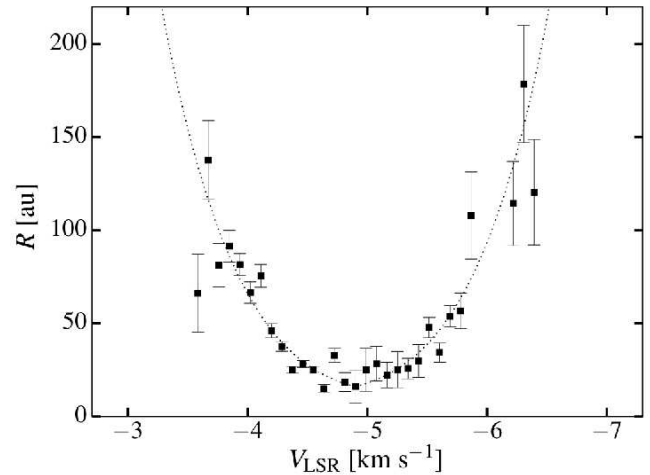}
\caption{Distance--velocity diagram of the CH$_{3}$OH masers. A dotted line indicates the best-fit result of our parabolic infall model (see section \ref{sec:discus:model}). 
The $x$ and $y$ axes show LOS velocities and a projected distances from the dynamical center ($x_{0}$, $y_{0}$), which were determined by the model-fitting, respectively. 
\label{fig:8}}
\end{figure}

Figure \ref{fig:8} presents a "distance--velocity" diagram that shows the relation between LOS velocities and projected distances from the averaged position of the maser distribution. 
The red-shifted and blue-shifted sides are clearly symmetrical about the systemic velocity. 
This fact strongly suggests that the LOS velocities are expressed as a parabolic function of a projected distance. 

The maser distribution in Figure \ref{fig:4} appears spiral-like rather than ring-like or linear; however, 
the parabolic velocity field cannot be explained by a simple rotating disk with a spiral arm. 
Three-dimensional structure is clearly required rather than a planar disk-like system, given the nearly face-on geometry. 
Alternatively, such a non-axisymmetric structure evokes the infall streams found in several low-mass protostars and protobinaries \citep[e.g.,][]{2010Sci...327..306M, 2013Natur.493..191C, 2014ApJ...793...10T, 2014ApJ...793....1Y}. 
In particular, recent ALMA observations toward a low-mass class 0 object have found that 
similar non-axisymmetric streams fall down onto the edge-on accretion disk along a parabolic orbit \citep{2014ApJ...793....1Y}. 

If such a system is observed in a face-on view, the LOS component of the infall velocity approaches zero near the center.  
On the other hand, increasing LOS components at outer radii can cancel a decrease of the net infall velocity along a radial distance. 
These behaviors consistently explain the overall profile of our distance--velocity diagram, hence, we have performed a model-fitting in order to examine such a model.

\subsection{Parabolic infall model}\label{sec:discus:model}
\begin{figure}[tb]
\epsscale{1.0}
\plotone{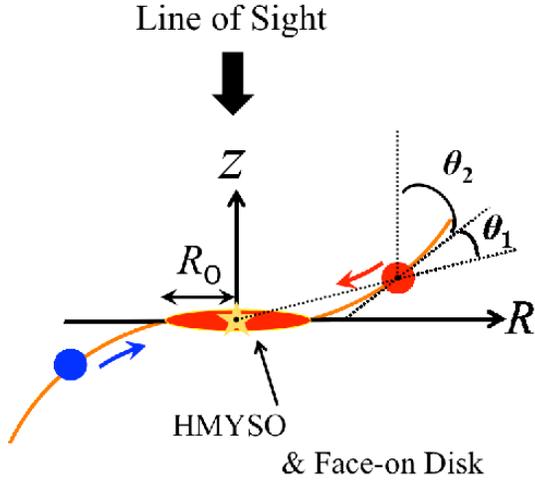}
\caption{Schematic view of our parabolic infall model. 
The blue and red filled circles indicate a blue-shifted and red-shifted infalling maser clumps, respectively. 
The LOS is along the $Z$ direction. 
$\theta_{1}$ and $\theta_{2}$, are angles between the direction of the dynamical center, tangential to the direction of the trajectory and LOS (see section \ref{sec:discus:model}). 
\label{fig:9}}
\end{figure}

Figure \ref{fig:9} shows the schematic view of our parabolic infall model. 
Here, infall streams traced by the CH$_{3}$OH masers fall down to the equatorial plane of the inner accretion disk, along the point symmetric trajectory. 
We assume a completely face-on system for simplicity. 
This allows us to ignore any rotating motion that is perpendicular to the LOS. 
We thus ignore any azimuthal coordinate, while the streamline may fall down along a 3D paraboloid in practice. 

The height from the equatorial plane $Z$, is modeled by the power-law function of radial distance $R$, analogous to the standard disk \citep[e.g.,][]{2011ApJ...733...55Z}: 
\begin{equation}
Z = Z_0 \left(\frac{R - R_{\rm 0}}{R_{\rm 0}}\right)^p,
\end{equation}
where $Z_0$, $p$, and $R_{\rm 0}$ indicate a scaling factor, power-law index, and landing radius where the infall streams reach the equatorial plane, respectively. 

The LOS velocity $V_{\rm mod}$ is modeled by the freefall velocity, assuming that only a tangential component along the streamline is always effective: 
\begin{equation}
V_{\rm mod} = V_{\rm sys} \pm \sqrt{\frac{2GM_{*}}{R^2 + Z^2}}\rm{cos}\theta_1\rm{cos}\theta_2, 
\end{equation}
where, $G$ and $M_{*}$ are the gravitational constant and central YSO mass, respectively. 
$\theta_1$ is an angle between the direction toward the gravitational center and the tangential direction of the model trajectory. 
$\theta_2$ is an angle between the tangential direction and the LOS (see Figure \ref{fig:9}). 
The sign in the equation was set as positive for the red-shifted stream and negative for the blue-shifted stream.  
$M_{*}$ was fixed at 10 $M_{\sun}$ as expected from the $L_{\rm bol}$. 
Although this may still be a rough estimation, it should be emphasized that our model is robust against $M_{*}$. 
We can always find a proper $Z_{0}$ for different $M_{*}$, i.e., larger $M_{*}$ simply results in lower $Z_{0}$. 
We also note that both $\theta_1$ and $\theta_2$ are automatically determined for each parameter set as a function of $R$.  

In addition to  $Z_0$, $R_{\rm 0}$ and $p$, 
we introduced three other parameters, i.e., the relative position of the dynamical center ($x_{0}$, $y_{0}$) with respect to the center of the maser distribution, and true systemic velocity $V_{\rm sys}$. 
We have performed reduced chi-squared fitting on the distance--velocity diagram using these six free parameters. 
Observed LOS velocities $V_{\rm obs}$ were fit by $V_{\rm mod}$ as minimizing $\chi^2$, 
\begin{equation} 
\chi^2 = \frac{1}{N}\sum_{i}\frac{(V^{\rm obs}_{i} - V^{\rm mod}_{i})^2}{\delta V^2}. 
\end{equation} 
$N$ and $\delta V$ are the degree of freedom ($N$ = 24) and error of observed LOS velocities. 
We simply adopt the spectral resolution of 0.0878 km s$^{-1}$ as $\delta V$.  

\begin{figure*}[!htb]
\epsscale{1.15}
\plotone{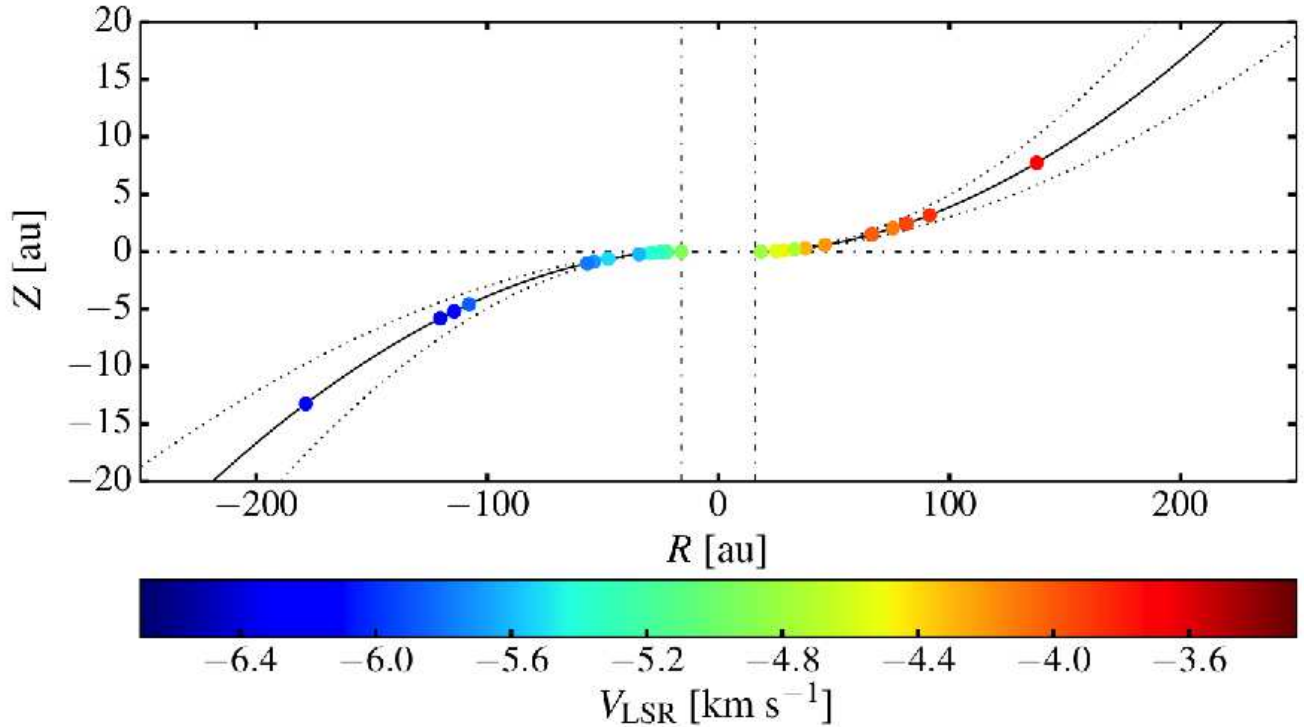}
\caption{Edge-on cross-sectional ($R$ - $Z$) view of the parabolic trajectory, i.e., the LOS is along the $Z$ direction. 
The solid and two dotted lines correspond to the best-fit model and 1 $\sigma$ error range, respectively. 
The positive and negative radial distances indicate an east blue-shifted region and west red-shifted region, respectively. 
The vertical and horizontal dash-dotted lines show the landing radius ($R$ = $\pm$ $R_{0}$) and equatorial plane ($Z$ = 0), respectively. 
The radial locations of observed data points are also presented by the filled circles, with the color indicating LOS velocities, where any azimuthal locations were ignored. 
\label{fig:10}}
\end{figure*}

The fitting results are shown in Figure \ref{fig:8} and the best-fit parameters are listed in Table \ref{tbl-8}. 
The error for each parameter was determined as $\chi^2$ to be unity. 
The minimized $\chi^2$ of $\sim$ 0.67 indicates that our parabolic infall model consistently explains the observed velocity field. 
Figure \ref{fig:10} shows the edge-on cross-sectional view of the best-fit trajectory with a 1 $\sigma$ error range. 
Positive and negative radial distances indicate the east blue-shifted region and west red-shifted region, respectively. 
Figure \ref{fig:10} also presents the radial locations of the observed data points along the best-fit streamline as a reference, where any azimuthal locations were ignored.

The streamline is quasi-radial ($Z$/$R$ $<$ 10 per cent) even at the outer region ($R$ $\sim$ 200 au). 
This fact indicates that centrifugal force is basically negligible, 
although the head--tail like distribution of the blue-shifted masers in Figure \ref{fig:4} may imply non-zero angular momentum. 
Such an infall-dominated motion of the class II CH$_{3}$OH maser at 100 au scale has actually been detected in a VLBI study by \citet{2011A&A...535L...8G}. 

The $R_{0}$ of 16 au is clearly smaller than that of typical accretion disks in High-Mass star-formation \citep[e.g., ][and references therein]{2016A&ARv..24....6B}. 
This fact indicates that the initial angular momentum in G353 is very small, 
or the CH$_{3}$OH masers selectively trace accreting material that has a small angular momentum. 
This point will be discussed again in section \ref{sec:discus:interp}

Another interesting feature is that the innermost maser distribution looks like a semi-ring as seen in Figure \ref{fig:4}. 
It is noteworthy that the radius and center of this semi-ring, roughly coincide with the best-fit landing radius and dynamical center. 
This structure may have some physical origin such as an accretion shock at the centrifugal barrier (i.e., a half of the centrifugal radius), as found in recent ALMA observations toward several low-mass class 0 objects \citep[e.g.,][]{2014Natur.507...78S, 2016ApJ...820L..34S, 2016ApJ...824...88O}. 

\begin{deluxetable*}{cccccccc}
\tabletypesize{\scriptsize}
\tablecaption{The best-fit parameters for the parabolic infall model \label{tbl-8}}
\tablewidth{0pt}
\tablehead{
\colhead{Reduced} & \colhead{$N$} & \colhead{$p$} & \colhead{$Z_{0}$} &\colhead{$R_{0}$}  & \colhead{$x_{0}$}&\colhead{$y_{0}$} & \colhead{$V_{\rm sys}$}\\
\colhead{-$\chi^{2}$} & \colhead{} & \colhead{} & \multicolumn2c{(au)} &  \multicolumn2c{(mas)} & \colhead{(km s$^{-1}$)}
} 
\startdata
0.68 & 24 &1.86 $\pm$ 0.07 & 0.18 $\pm$ 0.03 & 16.0 $^{+1.0}_{-0.9}$ & 2.3 $^{+4.9}_{-4.4}$ & 4.6 $^{+4.9}_{-5.5}$ & -4.9 $\pm$ 0.1\\
\enddata
\tablecomments{
The definition of each parameter is described in the main text. The formal errors were determined as $\chi^2$ to be unity. }
\end{deluxetable*}

\subsection{Physical interpretation}\label{sec:discus:interp}
Our infall model successfully explains the observed velocity structure, but the origin of that structure is still an open question. 
The spiral-like (or head--tail) azimuthal structure could be a natural outcome of gravitational acceleration with non-zero angular momentum. 
Gravitational interaction between binary companions and the disk can also cause such a streamline from circumbinary disk \citep[e.g.,][]{2012ApJ...749..118S}. 
It may be caused by gravitational instability in a self-gravitating disk \citep[e.g.,][]{2016Natur.538..483T}. 
Since the masers probably trace only part of the entire structure, a spatially resolved thermal line and/or continuum image is required. 

On the other hand, the power-law index $p$ of 1.86 is clearly larger than that for a scale height of a typical standard disk \citep[$p$ $\sim$ 1;][]{2011ApJ...733...55Z}, 
and is even steeper than that for isothermal disks \citep[$p$ $\sim$ 1.5;][]{1997ApJ...486..372B}. 
One possibility is that it is determined by the 2D profile of temperature and density, 
since the pumping condition of the class II CH$_{3}$OH maser emission is sensitive to these parameters. 

If the streamline traces the actual curvature of a circumstellar structure, a physically thick envelope (or toroid) may be more adequate. 
Alternatively, it can simply trace a trajectory of channel flows between hierarchical accretion systems, as in the case of low-mass objects \citep[e.g,][]{2013Natur.493..191C, 2014ApJ...793...10T}. 
Such a specific structure should be directly detected by a higher-resolution image of thermal dust or molecular emission. 
This may have already been indicated by the fine structure of our VLA image that is discussed below.

Another exotic explanation is that the infall trajectory is frozen in a hourglass-shaped magnetic field 
that is supposed to be common in both high- and low-mass star-formation \citep[e.g.,][]{2006Sci...313..812G, 2009Sci...324.1408G}. 
Maser polarization would be a direct diagnostic in this case \citep[e.g.,][]{2010MNRAS.404..134V, 2015A&A...583L...3S}. 
 
\begin{figure}[!tb]
\epsscale{1.2}
\plotone{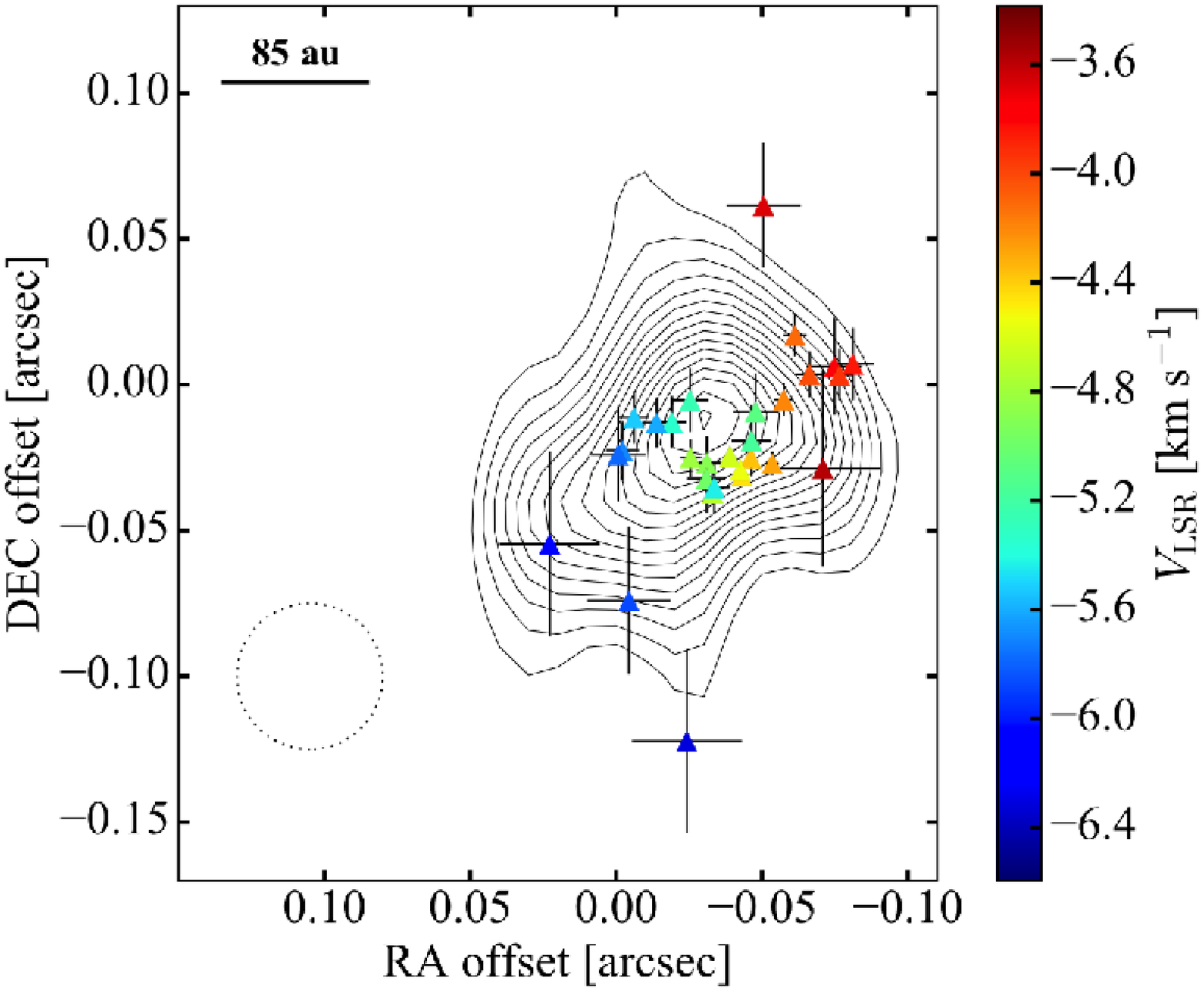}
\caption{"Super-resolution" image of the 45 GHz continuum that was constructed by all the clean components extracted in the beam deconvolution process. 
The contours are from 24\% ($3 \sigma$) to 99\%, with a step of 5\% of the image peak flux (1.13 mJy beam$^{-1}$). 
A circular beam of 50 mas diameter was convolved again, instead of the original synthesized beam. 
The adopted circular beam is shown in the lower left corner. 
The filled triangles and colors are the same as those in Figure \ref{fig:3}. \label{fig:11}}
\end{figure}

If the compact continuum emission is really explained by the optically thick dust continuum emission as proposed by our SED analysis in section \ref{sec:discus:cont}, 
a simple interpretation is that both of the continuum and masers trace the same dusty infall streams. 
Such a direct connection may be suggested by the "super-resolution" continuum image in Figure \ref{fig:11}. 
Here, all clean components extracted in the beam deconvolution process are reconvolved with a circular beam of 50 mas, 
which is the same as the minor axis of the original synthesized beam. 
The image shows that the dust continuum emission is clearly elongated along the maser distribution. 

A true rotationally supported accretion disk, if it exists, is expected to be smaller than 32 au in diameter (i.e., 2$R_{\rm 0}$) in this case. 
This is one of the most compact disks of the currently known accretion systems in high-mass star-formation \citep[e.g.,][]{2010Natur.466..339K}.  
If a centrifugal barrier is really located at 16 au in radius, the initial specific angular momentum in G353 is estimated to be $\sim$ 8 $\times$ 10$^{20}$ ($M_{*}$$/$10 $M_{\sun}$)$^{0.5}$ cm$^{2}$ s$^{-1}$. 
We note that the estimated angular momentum varies by less than 40\% within a YSO mass range of 5 -- 20 $M_{\sun}$. 
Another possible case is that a significant fraction of total angular momentum could be removed outside of 100 au. 
Although the centrifugal barrier is relatively small, the specific angular momentum expected in the former case is slightly larger than that of several low-mass disks and/or envelopes at the 100 au scale \citep[e.g.,][]{2017ApJ...834..178Y}, due to the larger stellar mass. 

All these hypotheses will be verified by thermal continuum and line observations at higher resolution ($\sim$ 10 mas), resolving a face-on disk, infall streams, and binary companions, etc. 
This will be done by our long-baseline project in ALMA cycle 4. 
In addition, our infall model itself will be directly examined by ongoing proper motion measurements of the class II CH$_{3}$OH masers by the Very Long Baseline Array (VLBA). 
This will allow us to independently determine $Z_{0}$ and $M_{*}$. 
A 3D velocity field will also constrain the accretion rate within 100 au, combined with mass information obtained by the ALMA. 

\section{Conclusions}\label{sec:concl}
We have performed J-VLA and ATCA observations, searching for the accretion system in G353 that is the best candidate for a face-on HMYSO. 
Detailed observational outcomes and outstanding implications are as follows.

\begin{enumerate} 
\item The bolometric luminosity of 5$\times$10$^{3}$ $L_{\sun}$ was estimated based on the IR archival data. This implies that the host YSO mass is around 10 $M_{\sun}$. 
\item Our SED analysis suggested that the overall centimeter SED could be modeled by a combination of the radio jet and dust continuum emission. 
 In particular, the compact 45 GHz continuum source detected by J-VLA could be explained by an optically thick dust emission from the innermost circumstellar system of 100 au radius, although any unresolved free-free contamination is still possible. 
 \item  The expected mass of the dusty system is 0.2 $M_{\sun}$ at minimum and up to a few $M_{\sun}$ depending on the dust parameters. 
 This may suggest that the system could be self-gravitating and dynamically unstable; however, the SED model is still tentative and spatially resolved SED measurements are required to confirm the exact circumstellar environment. 
 \item The associated class II CH$_{3}$OH masers have shown a spiral-like distribution. 
 The observed systematic velocity gradient is inconsistent with the Keplerian rotation and simple infall motion that are applicable for several class II CH$_{3}$OH maser sources. 
 We have alternatively found that it can be explained by infall streams falling down onto the face-on equatorial plane along a parabolic orbit. 
\item Our parabolic infall model predicts that the streamline is quasi-radial and reaches the equatorial plane at 16 au radius. 
This radius is clearly smaller than that of typical accretion disks in high-mmass star formation, indicating that initial angular momentum in G353 was very small, 
or the CH$_{3}$OH masers selectively trace accreting materials that have small angular momentum. 
\item The physical origin of such a streamline is still an open question and there are several possible scenarios. 
The spiral-like azimuthal distribution can be explained by some gravitational effects such as gravitational acceleration, binary interaction, and gravitational instability, etc. 
On the other hand, the vertical trajectory may be determined by (1) a 2D profile of temperature and density that is adequate for maser excitation, or 
(2) the actual shape of a physically thick envelope or channel flows between hierarchical accretion system, 
or (3) an hourglass-shaped magnetic field. 
\item The super-resolution image suggests that both of the continuum emission and masers trace the same dusty infall stream. 
In this case, the initial specific angular momentum in G353 is $\sim$ 8 $\times$ 10$^{20}$ ($M_{*}$$/$10 $M_{\sun}$)$^{0.5}$ cm$^{2}$ s$^{-1}$, 
or a significant fraction of initial angular momentum could be removed outside of 100 au. 
\end{enumerate}

Our pilot imaging successfully revealed the first face-on view of the innermost accretion system in high-mass star-formation, although it was still partially resolved. 
In the future, we will perform the highest-resolution ($\sim$ 10 mas) thermal continuum and line observations with the ALMA long baselines, 
in order to verify all these hypotheses and constrain the physical environment of the innermost region ($<$ 100 au). 
Proper motion measurements of the class II CH$_{3}$OH masers using VLBA are also ongoing. 
This will give a direct examination of our infall model itself, based on 3D maser kinematics.

\acknowledgments
We would like to thank the unknown referee for important suggestions on the initial draft of this paper.
The Jansky-VLA is operated by the National Radio Astronomy Observatory (NRAO). 
The NRAO is a facility of the National Science Foundation operated under cooperative agreement by Associated Universities, Inc. 
The Australia Telescope Compact Array is part of the Australia Telescope National Facility, which is funded by the Commonwealth of Australia for operation as a National Facility managed by CSIRO. 
This work was financially supported by a Grant-in-Aid from the Japan Society for the Promotion of Science Fellows (K.M.) 
and Grants-in-Aid by the Ministry of Education, Culture, and Science of Japan 24-6525 and 15K17613 (K.M.). 

This research has made use of the NASA/ IPAC Infrared Science Archive, which is operated by the Jet Propulsion Laboratory, California Institute of Technology, under contract with the National Aeronautics and Space Administration.
This research has also made use of data products from the Two Micron All Sky Survey, which is a joint project of the University of Massachusetts and the Infrared Processing and Analysis Center/California Institute of Technology, funded by the National Aeronautics and Space Administration and the National Science Foundation. 
This work is based, in part, on observations made with the Spitzer Space Telescope, which is operated by the Jet Propulsion Laboratory, California Institute of Technology under a contract with NASA.
This publication made use of data products from the Midcourse Space Experiment. Processing of the data was funded by the Ballistic Missile Defense Organization with additional support from NASA Office of Space Science. 
 The ATLASGAL project is a collaboration between the Max-Planck-Gesellschaft, the European Southern Observatory (ESO) and the Universidad de Chile. It includes projects E-181.C-0885, E-078.F-9040(A), M-079.C-9501(A), M-081.C-9501(A) plus Chilean data.
{\it Herschel} is an ESA space observatory with science instruments provided by European-led Principal Investigator consortia and with important participation from NASA.

\software{CASA \citep{2007ASPC..376..127M}, 
AIPS \citep{2003ASSL..285..109G}, 
MIRIAD \citep{1995ASPC...77..433S}, 
PyFITS \citep{1999ASPC..172..483B}, 
Matplotlib \citep{2007CSE.....9...90H}, 
} 

\facilities{VLA, ATCA, CTIO:2MASS, {\it Spitzer}, MSX, {\it Herschel}, APEX}

\end{document}